\documentclass[12pt]{article}
\usepackage{epsfig,amsmath,euscript}

\def\delslash{\rlap{\hspace{0.02cm}/}{\partial}}
\def\nslash{\rlap{\hspace{0.02cm}/}{n}}
\def\nbslash{\rlap{\hspace{0.02cm}/}{\bar n}}
\def\vslash{\rlap{\hspace{0.02cm}/}{v}}
\def\Dslash{\rlap{\hspace{0.07cm}/}{D}}
\def\calAslash{\rlap{\hspace{0.08cm}/}{{\EuScript A}}}
\def\calDslash{\rlap{\hspace{0.1cm}/}{{\EuScript D}}}

\def\A{{\EuScript A}}
\def\D{{\EuScript D}}
\def\X{{\EuScript X}}

\def\bm#1{\mbox{\boldmath$#1$\unboldmath}}

\begin{document}

\begin{titlepage}

\begin{flushright}
CLNS~04/1890\\
{\tt hep-ph/0409115}\\[0.2cm]
September 9, 2004
% v4: December 21, 2004
\end{flushright}

\vspace{0.7cm}
\begin{center}
\Large\bf 
Subleading Shape Functions in Inclusive B Decays
\end{center}

\vspace{0.8cm}
\begin{center}
{\sc Stefan W. Bosch${}^a$, Matthias Neubert${}^{a,b}$, and Gil Paz${}^a$}\\
\vspace{0.7cm}
{\sl ${}^a$\,Institute for High-Energy Phenomenology\\
Newman Laboratory for Elementary-Particle Physics, Cornell University\\
Ithaca, NY 14853, U.S.A.\\[0.3cm]
${}^b$\,School of Natural Sciences, Institute for Advanced Study\\
Princeton, NJ 08540, U.S.A.}
\end{center}

\vspace{1.0cm}
\begin{abstract}
\vspace{0.2cm}\noindent
The contributions of subleading shape functions to inclusive decay 
distributions of $B$ mesons are derived from a systematic two-step matching of 
QCD current correlators onto soft-collinear and heavy-quark effective theory. 
At tree-level, the results can be expressed in terms of forward matrix 
elements of bi-local light-cone operators. Four-quark operators, which arise 
at $O(\alpha_s)$, are included. Our results are in disagreement with some 
previous studies of subleading shape-function effects. A numerical analysis of 
$\bar B\to X_u\,l^-\bar\nu$ decay distributions suggests that power 
corrections are small, with the possible exception of the endpoint region of 
the charged-lepton energy spectrum.
\end{abstract}
\vfil

\end{titlepage}

\section{Introduction}

Inclusive decays of $B$ mesons into final states containing light particles, 
such as $\bar B\to X_u\,l^-\bar\nu$ and $\bar B\to X_s\gamma$, play a 
prominent role in the extraction of the Cabbibo-Kobayashi-Maskawa (CKM) 
matrix element $|V_{ub}|$ and in searches for physics beyond the Standard 
Model. Necessary experimental cuts in the analysis of these processes restrict 
the hadronic final state to have large energy, $E_X\sim m_{B}$, but only 
moderate invariant mass, $M_X\sim\sqrt{m_B\Lambda_{\rm QCD}}$. In this region 
of phase space, the inclusive rates can be calculated using a twist expansion, 
which resums infinite sets of power corrections into non-perturbative shape 
functions \cite{Neubert:1993ch,Bigi:1993ex,Mannel:1994pm}. 

The impressive performance of the BaBar, Belle, and CLEO experiments demands a 
continuous effort to reduce the theoretical uncertainties affecting 
extractions of $|V_{ub}|$ using inclusive $B$ decays. Recently, significant 
progress has been made by systematically incorporating higher-order 
perturbative corrections \cite{Bosch:2004th,Bauer:2003pi,Neubert:2004dd}. A 
careful estimate of the theoretical uncertainty on $|V_{ub}|$, extracted using 
a cut on $P_+=E_X-|\bm{P}_X|$, finds an error of 7\% due to variations of the 
leading shape function \cite{Bosch:2004bt}, which can be reduced significantly 
using information on the $\bar B\to X_s\gamma$ photon spectrum. With a 5\% 
relative theoretical error on $|V_{ub}|$, corrections suppressed by a power of 
$\Lambda_{\rm QCD}/m_b$ are considered the second largest source of 
uncertainty. The present paper is devoted to a more thorough study of power 
corrections to inclusive $B$-meson decay distributions in the shape-function 
region, using the two-step matching procedure developed in 
\cite{Bosch:2004th,Bauer:2003pi,Neubert:2004dd}. Subleading shape functions 
have been investigated first by Bauer, Luke, and Mannel \cite{Bauer:2001mh}, 
and their effects on various inclusive spectra have been analyzed by several 
groups \cite{Leibovich:2002ys,Bauer:2002yu,Neubert:2002yx,Burrell:2003cf}. We 
disagree with the findings of \cite{Bauer:2001mh,Bauer:2002yu} in some 
important aspects.

The hadronic physics in the inclusive semileptonic decay 
$\bar B\to X_u\,l^-\bar\nu$ is encoded in a hadronic tensor $W^{\mu\nu}$ 
defined via the discontinuity of the forward $B$-meson matrix element of a 
correlator of two flavor-changing weak currents 
$J^\mu=\bar u\,\gamma^\mu(1-\gamma_5)\,b$. We define
\begin{equation}
   W^{\mu\nu} = \frac{1}{\pi}\,\mbox{Im}\,
    \frac{\langle\bar B(v)|\,T^{\mu\nu}\,|\bar B(v)\rangle}{2 m_B} \,, \qquad
   T^{\mu\nu} = i \int d^4x\,e^{iq\cdot x}\,
    T\{ J^{\dagger\mu}(0)\,J^\nu(x) \} \,,
\end{equation}
where $v$ is the $B$-meson velocity and $q$ the momentum carried by the lepton 
pair. To be slightly more general, we will consider QCD currents of the type 
$J_i^\dagger=\bar b\,\Gamma_i\,q$ and $J_j=\bar q\,\Gamma_j\,b$, where $q$ is 
a massless quark, and $\Gamma_i$, $\Gamma_j$ are arbitrary Dirac matrices. The 
corresponding current correlator $T_{ij}$ and 
hadronic tensor $W_{ij}$ can then also be applied to study the 
contribution of the dipole operator $Q_{7\gamma}$ to $\bar B\to X_s\gamma$ 
decay.

The current correlator $T_{ij}$ receives contributions from different energy 
scales, which can be disentangled using effective field theories. Besides the 
hard scale $m_b$ and the hadronic scale $\Lambda_{\rm QCD}$, an intermediate 
``hard-collinear'' scale $\sqrt{m_b\Lambda_{\rm QCD}}$ set by the typical 
invariant mass of the hadronic final state is of relevance. Because quantum 
fluctuations associated with the hard and hard-collinear scales can be treated 
in perturbation theory, the hadronic tensor trivially factorizes into products 
of hard functions $(\mu\sim m_b$), jet functions 
$(\mu\sim\sqrt{m_b\Lambda_{\rm QCD}}$), and shape functions defined in terms 
of $B$-meson matrix elements of quark-gluon operators. In a first step, hard 
fluctuations are integrated out by matching the QCD currents onto 
soft-collinear effective theory (SCET) \cite{Bauer:2000yr,Beneke:2002ph}. 
The current correlator is then expanded in terms of light-cone 
operators in heavy-quark effective theory (HQET) \cite{Neubert:1993mb}, 
thereby integrating out fluctuations at the hard-collinear scale. The fact 
that the relevant HQET operators live on the light cone follows from the 
structure of the multipole expansion of soft fields in SCET. In the present 
work, we carry out the matching procedure at tree level and to order 
$\Lambda_{\rm QCD}/m_b$ in the heavy-quark expansion. We indicate how our 
results would change if loop corrections were included.

\section{Short-distance expansion of the hadronic tensor}

We begin by recalling some elements of SCET, referring to \cite{Beneke:2002ph} 
for a more detailed discussion (see also \cite{Bauer:2000yr}). We work in a 
reference frame where the $B$ meson is at rest, $v^\mu=(1,0,0,0)$, and where
the lepton 3-momentum $\bm{q}$ points in the negative $z$-direction. We
introduce two light-like vectors $n^\mu=(1,0,0,1)$ and 
$\bar n^\mu=(1,0,0,-1)$, with $n\cdot\bar n=2$, $n\cdot v=\bar n\cdot v=1$, 
and $\bar n=2v-n$. Any 4-vector $a^\mu$ can be decomposed as
\begin{equation}
   a^\mu = \bar n\cdot a\,\frac{n^\mu}{2} + n\cdot a\,\frac{\bar n^\mu}{2}
   + a_\perp^\mu \equiv a_-^\mu + a_+^\mu + a_\perp^\mu \,.
\end{equation}
In this basis, $v_\perp=0$ and $q_\perp=0$ by construction. In the kinematic 
region of interest, the hadronic jet has a hard-collinear momentum, which 
scales like $(P_-,P_+,P_\perp)\sim m_b(1,\lambda,\sqrt{\lambda})$, where 
$\lambda\sim\Lambda_{\rm QCD}/m_b$. The momenta of the soft, light 
constituents of the $B$ meson scale like 
$p_s^\mu\sim\Lambda_{\rm QCD}\sim m_b\lambda$. 

SCET is the appropriate effective field theory for the description of the 
interactions among soft and hard-collinear degrees of freedom. Its Lagrangian 
is organized in an expansion in powers of $\sqrt{\lambda}$. The leading-order 
Lagrangian is
\begin{eqnarray}\label{L0}
   {\cal L}_{\rm SCET}^{(0)}
   &=& \bar\xi\,\frac{\nbslash}{2}
    \left( in\cdot D_{hc} + g n\cdot A_s(x_-)
    + i\Dslash_{\perp hc}\,\frac{1}{i\bar n\cdot D_{hc}}\,i\Dslash_{\perp hc}
    \right) \xi \nonumber\\
   &&\mbox{}+ \bar q\,i\Dslash_s\,q + \bar h\,iv\cdot D_s\,h
    + {\cal L}_{\rm YM}^{(0)} \,,
\end{eqnarray}
where $\xi$ is a hard-collinear quark field, $q$ is a soft, massless quark 
field, $h$ is a heavy-quark field defined in HQET, $A_s$ is a soft gluon 
field, and $iD_{hc}^\mu=i\partial^\mu+g A_{hc}^\mu$ is the covariant 
derivative containing a hard-collinear gluon field. All fields in the above 
Lagrangian are evaluated at point $x$, except for the soft gluon field in the 
first term, which is evaluated at $x_-=\frac12\,(\bar n\cdot x)\,n$. The 
explicit form of the leading-order Yang-Mills Lagrangian can be found in 
\cite{Beneke:2002ph}. 

The terms up to second order in the expansion in $\sqrt{\lambda}$ are
\begin{eqnarray}
   {\cal L}_{\rm SCET}^{(1)}
   &=& {\cal L}_\xi^{(1)} + {\cal L}_{\xi q}^{(1)} + {\cal L}_{\rm YM}^{(1)}
    \,, \nonumber\\
   {\cal L}_{\rm SCET}^{(2)}
   &=& {\cal L}_\xi^{(2)} + {\cal L}_{\xi q}^{(2)} + {\cal L}_h^{(2)}
    + {\cal L}_{\rm YM}^{(2)} \,,
\end{eqnarray}
where
\begin{equation}
   {\cal L}_h^{(2)} = \frac{1}{2m_b} \left[ \bar h\,(iD_s)^2 h
   + \frac{C_{\rm mag}}{2}\,\bar h\,\sigma_{\mu\nu}\,g G_s^{\mu\nu} h
   \right]
\end{equation}
is the next-to-leading term in the expansion of the HQET Lagrangian 
\cite{Neubert:1993mb}. Expressions for the remaining Lagrangian corrections 
have been presented in \cite{Beneke:2002ph}.

While it is consistent to apply a perturbative expansion at the hard and 
hard-collinear scales, the soft-gluon couplings to hard-collinear fields are 
non-perturbative and must be treated to all orders in the coupling constant. 
For instance, the hard-collinear quark propagator derived from (\ref{L0}) 
should be taken to be the propagator in the background of the soft gluon 
field, summing up arbitrarily many insertions of the field $A_s$. The most 
convenient way of achieving this summation is to decouple the leading-order 
interactions between the soft gluon field and hard-collinear fields in 
(\ref{L0}) with the help of a field redefinition, under which 
\cite{Bauer:2000yr,Becher:2003qh}
\begin{equation}
   \xi(x) = S(x_-)\,\xi^{(0)}(x) \,, \qquad
   A_{hc}^\mu(x) = S(x_-)\,A_{hc}^{(0)\mu}(x)\,S^\dagger(x_-) \,,
\end{equation}
where
\begin{equation}
   S(x) = P\exp\Bigg( ig \int\limits_{-\infty}^0\!dt\,n\cdot A_s(x+tn)
   \Bigg)
\end{equation}
is a soft Wilson line along the $n$ direction. Introducing the new fields into 
the Lagrangian yields
\begin{equation}\label{Ldecoupled}
   {\cal L}_\xi^{(0)} = \bar\xi^{(0)}\,\frac{\nbslash}{2}
   \left( in\cdot D_{hc}^{(0)} 
   + i\Dslash_{\perp hc}^{(0)}\,\frac{1}{i\bar n\cdot D_{hc}^{(0)}}\,
   i\Dslash_{\perp hc}^{(0)} \right) \xi^{(0)} \,,
\end{equation}
and similarly all interactions between soft and hard-collinear gluon fields 
are removed from the Yang-Mills Lagrangian ${\cal L}_{\rm YM}^{(0)}$. The 
propagator of the new hard-collinear quark field is now given by the simple
expression
\begin{equation}\label{xiprop}
   \Delta_\xi(x-y) 
   = \langle 0|\,T\{ \xi^{(0)}(x)\,\bar\xi^{(0)}(y) \}\,|0\rangle
   = \frac{\nslash}{2} \int\frac{d^4p}{(2\pi)^4}\,
   \frac{i\bar n\cdot p}{p^2+i\epsilon}\,e^{-ip\cdot(x-y)} \,.
\end{equation}
The effect of soft-gluon attachments is taken into account by factors of the 
Wilson line $S$ in the results below.

We now list the expressions for the subleading corrections to the SCET
Lagrangian in terms of the redefined fields, using the formalism of 
gauge-invariant building blocks \cite{Hill:2002vw}. We define
\begin{equation}
   \X = W^\dagger \xi^{(0)} \,, \qquad
   \A_{hc}^\mu = W^\dagger (iD_{hc}^{(0)\mu} W) \,,
\end{equation}
where
\begin{equation}
   W = P\exp\Bigg( ig \int\limits_{-\infty}^0\!dt\,
   \bar n\cdot A_{hc}^{(0)}(x+t\bar n) \Bigg)
\end{equation}
is a hard-collinear Wilson line. These ``calligraphic'' fields are invariant
under both hard-collinear and soft gauge transformations. Note that in the
light-cone gauge, $\bar n\cdot A_{hc}^{(0)}=0$, we simply have $\X=\xi^{(0)}$ 
and $\A_{hc}^\mu=gA_{hc}^{(0)\mu}$. In terms of these fields, the results 
compiled in \cite{Beneke:2002ph} take the form
% split array for nicer page break (MN)
\begin{eqnarray}
   {\cal L}_\xi^{(1)} &=& \bar\X\,\frac{\nbslash}{2}\,x_\perp^\mu n^\nu
    \left( S^\dagger g G_{\mu\nu} S \right)_{x_-}\!\X \,, 
    \hspace{9.25cm} \nonumber
\end{eqnarray}
\begin{eqnarray}\label{L1and2}
   {\cal L}_\xi^{(2)} &=& \bar\X\,\frac{\nbslash}{2} \left( 
    \frac{n\cdot x}{2}\,\bar n^\mu n^\nu 
    \left( S^\dagger g G_{\mu\nu} S \right)_{x_-}
    + \frac{x_\perp^\mu x_\perp^\rho}{2}\,n^\nu
    \left( S^\dagger [D_\rho,g G_{\mu\nu}] S \right)_{x_-} \right)\!\X 
    \nonumber\\
   &+& \bar\X\,\frac{\nbslash}{2} \left( 
    i\calDslash_{\perp hc}\,\frac{1}{i\bar n\cdot\partial}\,
    \frac{x_\perp^\mu}{2}\,\gamma_\perp^\nu
    \left( S^\dagger g G_{\mu\nu} S \right)_{x_-}
    + \frac{x_\perp^\mu}{2}\,\gamma_\perp^\nu
    \left( S^\dagger g G_{\mu\nu} S \right)_{x_-}
    \frac{1}{i\bar n\cdot\partial}\,i\calDslash_{\perp hc} \right) \X \,,
    \nonumber\\
   {\cal L}_{\xi q}^{(1)} &=& \left( \bar q S \right)_{x_-}\!
    i\calDslash_{\perp hc}\,\X + \mbox{h.c.} \,,
\end{eqnarray}
where $i\D_{hc}^\mu=i\partial^\mu+\A_{hc}^\mu$, and we have dropped the 
subscript ``$s$'' on the soft covariant derivative and field strength. The 
expression for $ {\cal L}_{\xi q}^{(2)}$ will not be needed for our analysis. 
The notation $(\dots)_{x_-}$ indicates that, in interactions with 
hard-collinear fields, soft fields are multipole expanded and live at position 
$x_-$, whereas hard-collinear fields are always evaluated at position $x$. 
Because the SCET Lagrangian is not renormalized \cite{Beneke:2002ph}, the 
above expressions are valid to all orders in perturbation theory.

Next, we need the expressions for heavy-light current operators in SCET. In 
general, a QCD current $\bar q\,\Gamma\,b$ matches onto
\begin{equation}
   \bar q(x)\,\Gamma\,b(x)
   = e^{-im_b v\cdot x} \left( J_A^{(0)} + J_A^{(1)} + J_A^{(2)}
   + J_B^{(1)} + J_B^{(2)} + \dots \right) , 
\end{equation}
where we distinguish between type-$A$ ``two-particle'' operators and type-$B$ 
``three-particle'' operators \cite{Hill:2004if}. The operators arising at tree 
level are \cite{Beneke:2002ph,Hill:2004if,Pirjol:2002km}
\begin{eqnarray}
   J_A^{(0)} &=& \bar\X\,\Gamma \left( S^\dagger h \right)_{x_-} ,
    \nonumber\\
   J_A^{(1)} &=& \bar\X\,\Gamma\,x_\perp^\mu 
    \left( S^\dagger D_\mu h \right)_{x_-}
    + \bar\X\,\frac{\nbslash}{2}\,i\!\overleftarrow{\delslash}\!_\perp\,
    \frac{1}{i\bar n\cdot\overleftarrow{\partial}}\,\Gamma
    \left( S^\dagger h \right)_{x_-} , \nonumber\\
   J_A^{(2)} &=& \bar\X\,\Gamma \left[
    \frac{n\cdot x}{2} \left( S^\dagger \bar n\cdot D h \right)_{x_-} 
    + \frac{x_\perp^\mu x_\perp^\nu}{2}
    \left( S^\dagger D_\mu D_\nu h \right)_{x_-} 
    + \left( S^\dagger \frac{i\Dslash}{2m_b}\,h \right)_{x_-} \right]
    \nonumber\\
   &+& \bar\X\,\frac{\nbslash}{2}\,
    i\!\overleftarrow{\delslash}\!_\perp\,
    \frac{1}{i\bar n\cdot\overleftarrow{\partial}}\,\Gamma\,x_\perp^\mu
    \left( S^\dagger D_\mu h \right)_{x_-} ,
\end{eqnarray}
and
\begin{eqnarray}
   J_B^{(1)} &=&\mbox{}- \bar\X\,\frac{\nbslash}{2}\,\calAslash_{\perp hc}\,
    \frac{1}{i\bar n\cdot\overleftarrow{\partial}}\,\Gamma
    \left( S^\dagger h \right)_{x_-} 
    - \bar\X\,\Gamma\,\frac{\nslash}{2m_b}\,
    \calAslash_{\perp hc} \left( S^\dagger h \right)_{x_-} , \nonumber\\
   J_B^{(2)} &=&\mbox{}- \bar\X\,\Gamma \left(
    \frac{1}{i\bar n\cdot\partial} + \frac{\nslash}{2m_b} 
    \right) n\cdot\A_{hc} \left( S^\dagger h \right)_{x_-} \nonumber\\
   &&\mbox{}- \bar\X \left( \frac{\nbslash}{2}\,\calAslash_{\perp hc}\,
    \frac{1}{i\bar n\cdot\overleftarrow{\partial}}\,\Gamma
    + \Gamma\,\frac{\nslash}{2m_b}\,\calAslash_{\perp hc} \right)
    x_\perp^\mu \left( S^\dagger D_\mu h \right)_{x_-} \nonumber\\
   &&\mbox{}- \bar\X\,\Gamma\,\frac{1}{i\bar n\cdot\partial}\,
    \frac{(i\calDslash_{\perp hc}\,\calAslash_{\perp hc})}{m_b}
    \left( S^\dagger h \right)_{x_-}
    + \bar\X\,\frac{i\overleftarrow{\calDslash\!}_{\perp hc}}{m_b}\,
    \frac{1}{i\bar n\cdot\overleftarrow{\partial}}\,\frac{\nbslash}{2}\,
    \Gamma\,\frac{\nslash}{2}\,\calAslash_{\perp hc}
    \left( S^\dagger h \right)_{x_-} \,.
\end{eqnarray}
The expressions for the currents beyond tree level are more complicated, 
primarily because several new Dirac structures appear. The relevant formulae 
are known at leading \cite{Bauer:2000yr} and next-to-leading order 
\cite{Hill:2004if} in the power expansion. The corresponding results for the 
currents $ J_A^{(2)}$ and $J_B^{(2)}$ have not yet been derived. They would be 
needed if the analysis in this paper should be extended beyond tree level.

If perturbative corrections at the hard-collinear scale are neglected, the
hard-collinear gluon fields can be dropped, and the above expressions for the 
effective Lagrangians and currents simplify. In this approximation 
$\X\to\xi^{(0)}$, $\A_{hc}^\mu\to 0$, $i\D_{hc}^\mu\to i\partial^\mu$, and 
$J_B^{(n)}\to 0$. While this leads to great simplifications in the 
calculation, we stress that the structures of soft fields that arise do not 
simplify. We find operators containing $S^\dagger g G_{\mu\nu} S$, 
$S^\dagger [D_\rho,g G_{\mu\nu}] S$, $S^\dagger h$, $S^\dagger D_\mu h$, and 
$S^\dagger D_\mu D_\nu h$, and the same operators would arise if the 
calculation was extended beyond the tree approximation. The only exception is 
that we no longer retain Lagrangian corrections containing the soft quark 
field, because ${\cal L}_{\xi q}^{(1)}\to 0$ in the limit where the 
hard-collinear gluon field is neglected. (The term 
$\bar q S\,i\delslash_\perp\xi^{(0)}$ is forbidden by momentum conservation.) 
In Section~\ref{sec:4quark}, we analyze the subleading shape functions 
introduced at $O(\alpha_s)$ by two insertions of ${\cal L}_{\xi q}^{(1)}$.

\begin{figure}
\begin{center}
\epsfig{file=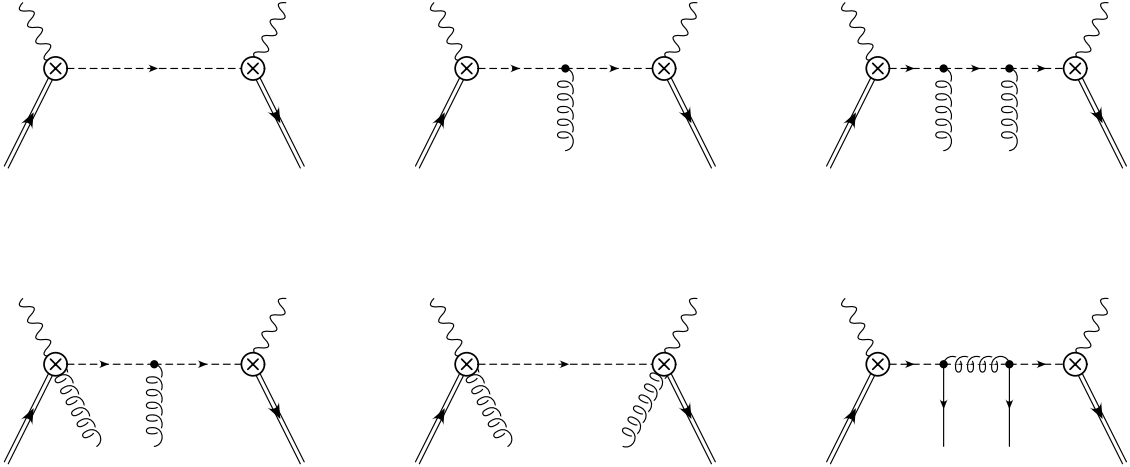,width=15cm}
\end{center}
\vspace{-0.2cm}
\centerline{\parbox{15cm}{\caption{\label{fig:graphs}
Representative examples of time-ordered products contributing to the current 
correlator $T_{ij}$ in SCET. Double lines show heavy-quark fields, dashed ones 
hard-collinear fields, and wavy lines denote the external currents. Lagrangian 
insertions and higher-order effective current operators are exemplified by 
soft gluons.}}}
\end{figure}

With all the definitions in place, we are now ready to evaluate the current
correlator $T_{ij}$ including terms of up to second order in $\sqrt{\lambda}$, 
working at lowest order in $\alpha_s$ at the hard and hard-collinear scales. 
The leading term is readily found to be
\begin{equation}\label{Tij0}
   T_{ij}^{(0)} = - \int d^4x\,e^{i(q-m_b v)\cdot x}
   \int \frac{d^4p}{(2\pi)^4}\,e^{ip\cdot x}\,
   \frac{\bar n\cdot p}{p^2+i\epsilon}\,
   \left( \bar h S \right)_{0} \Gamma_i\,\frac{\nslash}{2}\,\Gamma_j
   \left( S^\dagger h \right)_{x_-} .
\end{equation}
First-order corrections in $\sqrt{\lambda}$ vanish by rotational invariance in 
the transverse plane (provided we choose the coordinate system such that 
$v_\perp=0$ and $q_\perp=0$), i.e., $T_{ij}^{(1)}=0$. At second order in the 
expansion the correlator receives several contributions, which can be 
represented symbolically as
\begin{equation}\label{JJ}
   J^{\dagger(2)}\,J^{(0)} \,, \quad
   J^{\dagger(1)}\,J^{(1)} \,, \quad
   J^{\dagger(0)}\,J^{(2)} \,,
\end{equation}
\begin{equation}\label{JJL}
   J^{\dagger(1)}\,J^{(0)} \int d^4z\,{\cal L}_\xi^{(1)} \,,
    \quad
   J^{\dagger(0)}\,J^{(1)} \int d^4z\,{\cal L}_\xi^{(1)} \,,
    \quad
   J^{\dagger(0)}\,J^{(0)} \int d^4z \left[ {\cal L}_\xi^{(2)}
    + {\cal L}_h^{(2)} \right] ,
\end{equation}
and
\begin{equation}\label{JJLL}
   J^{\dagger(0)}\,J^{(0)} \int d^4z\,{\cal L}_\xi^{(1)}
   \int d^4w\,{\cal L}_\xi^{(1)} \,,
\end{equation}
where at tree level only the type-$A$ current operators appear. Examples of 
these time-ordered products are depicted in Figure~\ref{fig:graphs}. At 
$O(\alpha_s)$, one must include a tree-level contribution of the form
\begin{equation}\label{Lxiq2}
   J^{\dagger(0)}\,J^{(0)} \int d^4z\,{\cal L}_{\xi q}^{(1)}
   \int d^4w\,{\cal L}_{\xi q}^{(1)} \,,
\end{equation}
shown by the last diagram in the figure. Beyond tree level, one would also 
have to include the contributions from the type-$B$ current operators.

Because the external $B$-meson states in the definition of the hadronic tensor 
contain only soft constituents, and because the hard-collinear fields have 
been decoupled from the soft fields in the leading-order SCET Lagrangian 
(\ref{Ldecoupled}), it is possible to contract all hard-collinear fields in 
the time-ordered products. At tree level, all we need is the hard-collinear 
quark propagator (\ref{xiprop}). Derivatives acting on hard-collinear fields 
give powers of $p$ in momentum space, whereas components of $x^\mu$ appearing
in the multipole-expanded expressions for the effective Lagrangians and 
currents can be turned into derivatives $\partial/\partial p$ acting on the 
momentum-space amplitudes. Each insertion of a SCET Lagrangian correction 
introduces an integral over soft fields located along the $n$ light-cone. 
Consequently, the time-ordered products in (\ref{JJ}) lead to expressions 
involving bi-local operators as in (\ref{Tij0}), while those in (\ref{JJL}), 
(\ref{JJLL}), and (\ref{Lxiq2}) also lead to tri- and quadri-local operators.
At first sight, this would seem to require the introduction of complicated 
subleading shape functions depending on up to three momentum variables 
$\omega_i$, defined in terms of the Fourier transforms of the matrix elements 
of the non-local operators. However, the non-localities can be reduced using 
partial-fraction identities for the resulting hard-collinear quark 
propagators. At tree level, it suffices to define shape functions of a single 
variable $\omega$.

To see how this works, consider the effect of an insertion of the Lagrangian 
${\cal L}_\xi^{(1)}$ in (\ref{L1and2}). Since all hard-collinear fields must 
be contracted, we may consider without loss of generality the expression
\begin{equation}\label{typical}
   L(y_1,y_2) 
   = i\!\int d^4z\,\Delta_\xi(y_1-z)\,\frac{\nbslash}{2}\,z_\perp^\mu\,n^\nu
   \left( S^\dagger g G_{\mu\nu} S \right)_{z_-} \Delta_\xi(z-y_2) \,. 
\end{equation}
The fact that the soft fields live at position $z_-$ implies that the two 
hard-collinear quark propagators carry the same momentum components 
$\bar n\cdot p$ and $p_\perp$. In analogy with the definition of the 
hard-collinear calligraphic gluon field, we now introduce the soft field 
\cite{Hill:2002vw}
\begin{equation}\label{Asdef}
   \A_{s\mu}(x) = S^\dagger(x)\,(iD_\mu S)(x)
   = - \int\limits_{-\infty}^0\!dt\,n^\nu 
   \left( S^\dagger g G_{\mu\nu} S \right)(x+tn) \,,
\end{equation}
which allows us to write $n^\nu\,(S^\dagger g G_{\mu\nu} S)_{z_-}$ as a 
derivative, $-n\cdot\partial_z\,\A_{s\mu}(z_-)$. Integrating by parts in 
(\ref{typical}), we find
\begin{eqnarray}
   L(y_1,y_2) &=& i\!\int d^4z\,\Delta_\xi(y_1-z)\,
    \frac{\nbslash}{2}\,z_\perp^\mu\,\A_{s\mu}(z_-)\,
    n\cdot\partial_z\,\Delta_\xi(z-y_2) \nonumber\\
   &+& i\!\int d^4z\,[n\cdot\partial_z\,\Delta_\xi(y_1-z)]\,
    \frac{\nbslash}{2}\,z_\perp^\mu\,\A_{s\mu}(z_-)\,\Delta_\xi(z-y_2) . 
\end{eqnarray}
The hard-collinear propagator is a Green's function obeying the differential 
equation
\begin{equation}
   \left( n\cdot\partial + \frac{\partial_\perp^2}{\bar n\cdot\partial}
   \right) \Delta_\xi(x) = \frac{\nslash}{2}\,\delta^{(4)}(x) \,.
\end{equation}
This allows us to write
\begin{eqnarray}
   L(y_1,y_2)
   &=& i \Big[ y_{2\perp}\cdot\A_s(y_{2-}) - y_{1\perp}\cdot\A_s(y_{1-})
    \Big]\,\Delta_\xi(y_1-y_2) \nonumber\\
   &&\mbox{}- i\!\int d^4z\,\Delta_\xi(y_1-z)\,\frac{\nbslash}{2}\,
    z_\perp\cdot\A_s(z_-)\,
    \frac{\partial_{z\perp}^2}{\bar n\cdot\partial_z}\,\Delta_\xi(z-y_2)
    \nonumber\\
   &&\mbox{}- i\!\int d^4z
    \left[ \frac{\partial_{z\perp}^2}{\bar n\cdot\partial_z}\,
    \Delta_\xi(y_1-z) \right]
    \frac{\nbslash}{2}\,z_\perp\cdot\A_s(z_-)\,\Delta_\xi(z-y_2) . 
\end{eqnarray}
The terms involving transverse derivatives vanish at tree level, because they 
provide powers of transverse momenta of external lines, which are zero (recall 
that $v_\perp=0$ and $q_\perp=0$). This leaves the terms shown in the first 
line, in which the $z$ integral has been eliminated, and in which the product 
of two propagators has been reduced to a single propagator. In momentum space, 
these manipulations correspond to the partial-fraction identity
\begin{equation}\label{partialfractions}
   \frac{1}{\bar n\cdot p\,n\cdot p+p_\perp^2}\,
   \frac{1}{\bar n\cdot p\,(n\cdot p+\omega)+p_\perp^2}
   = \frac{1}{\bar n\cdot p\,\omega}
   \left[ \frac{1}{\bar n\cdot p\,n\cdot p+p_\perp^2}
   -    \frac{1}{\bar n\cdot p\,(n\cdot p+\omega)+p_\perp^2} \right] ,
\end{equation}
valid for two momenta that differ only in their $n\cdot p$ components.

In the sum of all terms many cancellations and simplifications take place, and
we find the rather simple result
\begin{equation}\label{Tresult}
   T_{ij}^{(2)} = - \int d^4x\,e^{i(q-m_b v)\cdot x}
   \int \frac{d^4p}{(2\pi)^4}\,e^{ip\cdot x}\,
   \frac{\bar n\cdot p}{p^2+i\epsilon}\,\sum_{n=1}^4 O_n(x_-) \,,
\end{equation}
where
\begin{eqnarray}
   O_1(x_-) &=& i\int d^4z\,T\{
    \left( \bar h S \right)_{0} \Gamma_i\,\frac{\nslash}{2}\,\Gamma_j 
    \left( S^\dagger h \right)_{x_-} {\cal L}_h^{(2)}(z) \} \,, \nonumber\\
   O_2(x_-)
   &=& \frac{1}{2m_b} \left[ 
    \left( \bar h S \right)_{0} \Gamma_i\,\frac{\nslash}{2}\,\Gamma_j 
    \left( S^\dagger i\Dslash h \right)_{x_-}
    + \left( \bar h (-i\!\overleftarrow{\Dslash}) S \right)_{0}
    \Gamma_i\,\frac{\nslash}{2}\,\Gamma_j \left( S^\dagger h \right)_{x_-} 
    \right] , \nonumber\\
   O_3(x_-)
   &=& \frac{1}{\bar n\cdot p} \left[
    \left( \bar h S \right)_{0} \Gamma_i\,\frac{\nslash\nbslash}{4}\,
    \gamma_\mu\,\Gamma_j \left( S^\dagger iD_\perp^\mu h \right)_{x_-}
    + \left( \bar h iD_\perp^\mu S \right)_{0} \Gamma_i\,\gamma_\mu\,
    \frac{\nbslash\nslash}{4}\,\Gamma_j \left( S^\dagger h \right)_{x_-}
    \right] , \nonumber\\[-0.15cm]
   O_4(x_-)
   &=& \frac{i}{\bar n\cdot p} \int\limits_0^{\bar n\cdot x/2}\!dt\,
    \left( \bar h S \right)_{0} \Gamma_i\,
    \Big( S^\dagger i\Dslash_\perp \frac{\nslash}{2}\,i\Dslash_\perp S
    \Big)_{tn}\,\Gamma_j \left( S^\dagger h \right)_{x_-} .
\end{eqnarray}
In deriving these expressions, we have made use of the identity 
$iD^\mu=S (i\partial^\mu+\A_s^\mu) S^\dagger$ with $\A_s^\mu$ as defined in 
(\ref{Asdef}). Note that on the space of forward matrix elements the result 
for $T_{ij}^{(2)}$ is hermitean, because we can integrate by parts and use 
translational invariance.

In the last step, we can simplify the operator $O_2$ by noting that the HQET
equation of motion, $iv\cdot D h=0$, along with $\vslash h=h$, implies
\begin{equation}
   S^\dagger i\Dslash h = S^\dagger i\Dslash_\perp h
   + (\vslash-\nslash)\,in\cdot\partial \big( S^\dagger h \big) \,.
\end{equation}
It follows that all gauge-covariant derivatives of the heavy-quark fields are
perpendicular derivatives. This fact restricts the number of subleading shape 
functions.

\section{Definition of subleading shape functions}

The Dirac structure of the operators $O_n$ can be simplified noting that the 
heavy-quark fields $h$ of HQET are two-component spinor fields, so that
between $\bar h\dots h$ the Dirac basis collapses to a set of four basis 
matrices $(\bm{1},\bm{\sigma})$. In four-component notation, these are the
upper-left $2\times 2$ blocks of $(\bm{1},\bm{\gamma}\gamma_5)$ (in the Dirac
representation). Instead of $\bm{\gamma}\gamma_5$, we are free to take the 
matrices $\gamma_\perp^\mu\gamma_5$ and $\nslash\gamma_5$. It follows that, 
between the $P_v=\frac12(1+\vslash)$ projectors supplied by the heavy-quark 
fields, any Dirac matrix $\Gamma$ can be decomposed as
\begin{equation}\label{HQETdecomp}
   \Gamma \to \frac12\,\mbox{tr} \left( \Gamma\,P_v \right) \bm{1}
   - \frac12\,\mbox{tr} \left[ \Gamma\,P_v\,(\vslash-\nslash)\,\gamma_5
   \right] \nslash\gamma_5
   - \frac12\,\mbox{tr} \left( \Gamma\,P_v\,\gamma_{\perp\mu}\gamma_5 \right) 
   \gamma_\perp^\mu\gamma_5 \,.
\end{equation}
We denote by
\begin{equation}
   \langle\bar h\dots h\rangle
   \equiv \frac{\langle\bar B(v)|\,\bar h\dots h\,|\bar B(v)\rangle}{2m_B}
\end{equation}
the forward $B$-meson matrix element of any HQET operator. Rotational 
invariance in the transverse plane implies that transverse indices can only be 
contracted using the symmetric and anti-symmetric tensors (we set 
$\epsilon_{0123}=1$)
\begin{equation}
   g_\perp^{\mu\nu} = g^{\mu\nu} - \frac{n^\mu\bar n^\nu+n^\nu\bar n^\mu}{2}
    \,, \qquad
   \epsilon_\perp^{\mu\nu}
   = \epsilon^{\mu\nu\alpha\beta}\,v_\alpha n_\beta \,.
\end{equation}
It follows that the only non-vanishing matrix elements are
\begin{eqnarray}\label{shapes}
   \langle \left(\bar h S \right)_{0} \left( S^\dagger h \right)_{x_-} \rangle
   &=& \int d\omega\,e^{-\frac{i}{2}\omega\bar n\cdot x}\,S(\omega) \,,
    \nonumber\\
   \langle i\int d^4z\,T\{ 
   \left(\bar h S \right)_{0} \left( S^\dagger h \right)_{x_-}
   {\cal L}_h^{(2)}(z) \} \rangle
   &=& \frac{1}{m_b} \int d\omega\,e^{-\frac{i}{2}\omega\bar n\cdot x}\,
    s(\omega) \,, \nonumber\\
   \langle \left(\bar h S \right)_{0} \gamma_\perp^\rho\gamma_5
    \left( S^\dagger iD_\perp^\mu h \right)_{x_-} \rangle
   &=& - \frac{i\epsilon_\perp^{\rho\mu}}{2} 
    \int d\omega\,e^{-\frac{i}{2}\omega\bar n\cdot x}\,t(\omega) \,,
    \nonumber\\[-0.15cm]
   -i\int\limits_0^{\bar n\cdot x/2}\!dt\,\langle \left(\bar h S \right)_{0}
    \left( S^\dagger iD_\perp^\mu iD_\perp^\nu S \right)_{tn}
    \left( S^\dagger h \right)_{x_-} \rangle
   &=& \frac{g_\perp^{\mu\nu}}{2}
    \int d\omega\,e^{-\frac{i}{2}\omega\bar n\cdot x}\,u(\omega) \,,
    \nonumber\\[-0.2cm]
   -i\int\limits_0^{\bar n\cdot x/2}\!dt\,
   \langle \left(\bar h S \right)_{0} \nslash\gamma_5
    \left( S^\dagger iD_\perp^\mu iD_\perp^\nu S \right)_{tn}
    \left( S^\dagger h \right)_{x_-} \rangle
   &=& - \frac{i\epsilon_\perp^{\mu\nu}}{2}
    \int d\omega\,e^{-\frac{i}{2}\omega\bar n\cdot x}\,v(\omega) \,.
\end{eqnarray}
If radiative corrections at the hard scale are included, it would be more
appropriate to split up 
$s(\omega)=s_{\rm kin}(\omega)+C_{\rm mag}\,s_{\rm mag}(\omega)$, where
$C_{\rm mag}$ is the Wilson coefficient of the chromo-magnetic operator in
the subleading HQET Lagrangian. This ensures that the shape functions remain
independent of the heavy-quark mass. The definitions of the functions $t$, 
$u$, $v$ are chosen such that
\begin{eqnarray}\label{tuvdefs}
   \langle\bar h(0)\,\nslash\,[0,x_-]\,(i\Dslash_\perp h)(x_-)\rangle
   &=& \int d\omega\,e^{-\frac{i}{2}\omega\bar n\cdot x}\,t(\omega) \,,
    \nonumber\\[-0.15cm]
   -i\int\limits_0^{\bar n\cdot x/2}\!dt\,
   \langle\bar h(0)\,[0,tn]\,(iD_\perp)^2(tn)\,[tn,x_-]\,h(x_-)\rangle
   &=& \int d\omega\,e^{-\frac{i}{2}\omega\bar n\cdot x}\,u(\omega) \,,
    \nonumber\\[-0.2cm]
   -i\int\limits_0^{\bar n\cdot x/2}\!dt\,
   \langle\bar h(0)\,\frac{\nslash}{2}\,[0,tn]\,
   \sigma_{\mu\nu}^\perp\,gG_\perp^{\mu\nu}(tn)\,[tn,x_-]\,h(x_-)\rangle 
   &=& \int d\omega\,e^{-\frac{i}{2}\omega\bar n\cdot x}\,v(\omega) \,,
\end{eqnarray}
where $[x,y]\equiv S(x)\,S^\dagger(y)$ is a product of two infinite-length 
soft Wilson lines, which on the light cone (i.e., for $x,y\parallel n$) 
collapses to a straight Wilson line of finite length connecting $x$ and $y$. 

We also need a variation of the first matrix element in (\ref{tuvdefs}), in
which the derivative is located at position 0. Using hermitean conjugation, 
translational invariance, and the reality of $t(\omega)$,  which follows from
parity and time-reversal invariance of the strong interactions, we find that
\begin{equation}
   \langle(\bar h i\Dslash_\perp)(0)\,\nslash\,[0,x_-]\,h(x_-)\rangle
   = \langle\bar h(0)\,\nslash\,[0,x_-]\,(i\Dslash_\perp h)(x_-)\rangle \,,
\end{equation}
implying that all terms containing a single insertion of $D_\perp$ can be 
related to the function $t(\omega)$. From this relation, it follows that
\begin{eqnarray}\label{talt}
   \int d\omega\,e^{-\frac{i}{2}\omega\bar n\cdot x}\,t(\omega)
   &=& \langle \left( \bar h S \right)_{0} \frac{\nslash}{2}\,
    \Big[ \calAslash_{s\perp}(x_-) - \calAslash_{s\perp}(0) \Big]
    \left( S^\dagger h \right)_{x_-} \rangle \nonumber\\
   &=& - \int\limits_0^{\bar n\cdot x/2}\!dt\,
    \langle\bar h(0)\,\frac{\nslash}{2}\,[0,tn]\,\gamma_\perp^\mu n^\nu 
    g G_{\mu\nu}(tn)\,[tn,x_-]\,h(x_-)\rangle \,,
\end{eqnarray}
which defines the function $t(\omega)$ in terms of a matrix element of the 
field-strength tensor.

It is now straightforward to express the forward matrix element of the 
current correlator $T_{ij}$ in terms of shape functions. The resulting traces
of Dirac matrices can be simplified using identities for the 
$\epsilon_\perp^{\mu\nu}$ tensor derived in \cite{Lange:2003pk}. Taking the 
imaginary part, we obtain for the hadronic tensor
\begin{eqnarray}\label{final}
   W_{ij}^{(0)}
   &=& \int d\omega\,\delta(n\cdot p+\omega)\,S(\omega)\,T_1 \,, \qquad
    W_{ij}^{(1)} = 0 \,, \\
   W_{ij}^{(2)} &=& \int d\omega\,\delta(n\cdot p+\omega) \left[
    \frac{\omega\,S(\omega) + t(\omega)}{m_b}\,T_2
    + \frac{s(\omega)}{m_b}\,T_1
    + \frac{t(\omega)}{\bar n\cdot p}\,T_3
    + \frac{u(\omega)}{\bar n\cdot p}\,T_1
    - \frac{v(\omega)}{\bar n\cdot p}\,T_4 \right] , \nonumber
\end{eqnarray}
where now $p=m_b v-q$, and
\begin{eqnarray}\label{traces}
   T_1 &=& \frac14\,\mbox{tr}\left[ \Gamma_i\,\nslash\,\Gamma_j\,
    \frac{1+\vslash}{2} \right] , \hspace{1.05cm}
   T_3 = \frac14\,\mbox{tr}\left[ \Gamma_i\,\gamma_\rho^\perp\gamma_5\,
    \Gamma_j\,\frac{1+\vslash}{2}\,\gamma_\perp^\rho\gamma_5 \right] ,
    \nonumber\\
   T_2 &=& \frac18\,\mbox{tr}\,\Big[ \Gamma_i\,\nslash\,\Gamma_j\,
    (\vslash-\nslash) \Big] \,, \qquad
   T_4 = \frac14\,\mbox{tr}\left[ \Gamma_i\,\nslash\gamma_5\,\Gamma_j\,
    \frac{1+\vslash}{2}\,(\vslash-\nslash)\,\gamma_5 \right] .
\end{eqnarray}
It follows from (\ref{final}) that the subleading shape functions $s(\omega)$
and $u(\omega)$ always come with the same trace as the leading shape function 
$S(\omega)$. However, $u(\omega)$ is divided by the kinematic variable 
$\bar n\cdot p$, and so it does {\em not\/} enter in a universal (i.e., 
process-independent) combination with $S(\omega)$.
 
We can now specialize our result to the case of semileptonic decay, for which 
$\Gamma_i=\gamma^\mu(1-\gamma_5)$ and $\Gamma_j=\gamma^\nu(1-\gamma_5)$. This 
yields
\begin{eqnarray}\label{Wsl}
   W^{\mu\nu}
   &=& \int d\omega\,\delta(n\cdot p+\omega)\,\Bigg\{\!
    \left( n^\mu v^\nu + n^\nu v^\mu - g^{\mu\nu}
    - i\epsilon^{\mu\nu\alpha\beta}\,n_\alpha v_\beta \right) \nonumber\\
   &&\times \left[ \left( 1 + \frac{\omega}{m_b} \right) S(\omega)
    + \frac{s(\omega) + t(\omega)}{m_b}
    + \frac{u(\omega) - v(\omega)}{\bar n\cdot p} \right] \nonumber\\
  &&\mbox{}- 2 (n^\mu v^\nu + n^\nu v^\mu)\,\frac{t(\omega)}{\bar n\cdot p}
   + 2 n^\mu n^\nu \left[ - \frac{\omega\,S(\omega)}{m_b}
   - \frac{t(\omega)}{m_b}
   + \frac{t(\omega) + v(\omega)}{\bar n\cdot p} \right] \Bigg\} \,.
\end{eqnarray}
Similarly, for the contribution of the dipole operator $Q_{7\gamma}$ to 
$\bar B\to X_s\gamma$ decay, the Dirac structures are (up to prefactors) 
$\Gamma_i=\frac14[\gamma_\perp^\mu,\nbslash](1-\gamma_5)$ and
$\Gamma_j=\frac14[\nbslash,\gamma_\perp^\nu](1+\gamma_5)$, where the indices
$\mu,\nu$ are contracted with the transverse polarization vector of the photon.
In this case we obtain
\begin{equation}\label{Wgamma}
   W^{\mu\nu} = (i\epsilon_\perp^{\mu\nu} - g_\perp^{\mu\nu})
   \int d\omega\,\delta(n\cdot p+\omega)
   \left[ \left( 1 - \frac{\omega}{m_b} \right) S(\omega)
   + \frac{s(\omega) - t(\omega)}{m_b}
   + \frac{u(\omega) - v(\omega)}{\bar n\cdot p} \right] .
\end{equation}
We stress, however, that while at leading power in $\Lambda_{\rm QCD}/m_b$ the 
dipole operator gives the only tree-level contribution to the 
$\bar B\to X_s\gamma$ decay rate, this is no longer the case when power 
corrections are included. For instance, interference terms of the dipole 
operator with current-current operators can lead to new subleading shape 
functions even at lowest order in perturbation theory. To derive these 
structures, it would be necessary to match the entire effective weak 
Hamiltonian for $\bar B\to X_s\gamma$ decay onto SCET operators 
\cite{Neubert:2004dd}. This task still has to be completed beyond the leading 
order in $\lambda$. Contrary to claims in \cite{Bauer:2001mh,Bauer:2002yu}, a 
complete description of tree-level subleading shape-function effects in 
$\bar B\to X_s\gamma$ decay is therefore still lacking.

\section{Moment relations and comparison with the literature}

Moments of the shape functions can be related to forward $B$-meson matrix 
elements of local HQET operators \cite{Neubert:1993ch}. In particular, setting 
$x=0$ in the defining relations (\ref{shapes}) yields expressions for the 
normalization integrals of the shape functions. They are
\begin{equation}\label{norms}
   \int d\omega\,S(\omega) = 1 \,, \qquad
   \int d\omega\,\{s(\omega), t(\omega), u(\omega), v(\omega)\} = 0 \,.
\end{equation}
The vanishing of the norm of all subleading shape functions is a consequence 
of Luke's theorem \cite{Luke:1990eg}, and it ensures that there are no 
first-order $\Lambda_{\rm QCD}/m_b$ corrections to total inclusive decay 
rates. For the functions $t$, $u$, $v$ this is an obvious consequence of the 
fact that the integration domain in (\ref{talt}) and (\ref{tuvdefs}) shrinks 
to zero in the limit $x\to 0$. The interpretation of (\ref{norms}) is that 
subleading shape functions lead to local distortions of inclusive spectra, 
which cancel out when the spectra are integrated over a sufficiently large 
region in phase space. The first moments characterize the strength of the 
distortions, while higher moments determine their shape.

Taking a derivative $in\cdot\partial_x$ in the definitions (\ref{shapes}) 
brings down a factor of $\omega$ under the integrals on the right-hand side. 
Setting then $x\to 0$ yields a set of relations for the first moments of the 
shape functions. The resulting matrix elements can be evaluated by means of 
the relations \cite{Falk:1992wt}
\begin{equation}
   \langle \bar h\,\Gamma_{\alpha\beta}\,iD^\alpha iD^\beta h\rangle
   = \frac12\,\mbox{tr} \left( \Gamma_{\alpha\beta}\,\frac{1+\vslash}{2}
   \left[ \left( g^{\alpha\beta} - v^\alpha v^\beta \right)
   \frac{\lambda_1}{3} + i\sigma^{\alpha\beta}\,\frac{\lambda_2}{2} \right]
   \frac{1+\vslash}{2} \right) , 
\end{equation} 
and \cite{Manohar:1993qn}
\begin{equation}
   \langle i\int d^4z\,T\{ (\bar h iD^\mu h)(0)\,{\cal L}_h^{(2)}(z) \}
   \rangle
   = - v^\mu\,\langle {\cal L}_h^{(2)}(0) \rangle
   =  - v^\mu\,\frac{\lambda_1 + 3C_{\rm mag}\,\lambda_2}{2m_b} \,, 
\end{equation} 
where $\lambda_1$ and $\lambda_2$ are the familiar HQET parameters arising in 
the parameterization of second-order power corrections to inclusive decay 
spectra, and $C_{\rm mag}=1$ at tree level. We obtain
\begin{eqnarray}\label{moments}
   \int d\omega\,\omega\,s(\omega)
   &=& - \frac{\lambda_1 + 3\lambda_2}{2} \,, \qquad 
   \int d\omega\,\omega\,u(\omega) = \frac{2\lambda_1}{3} \,, \nonumber\\
   \int d\omega\,\omega\,t(\omega)
   &=& - \lambda_2 \,, \hspace{2.05cm}
   \int d\omega\,\omega\,v(\omega) = \lambda_2 \,.
\end{eqnarray}
We also recall that the first two moments of the leading shape function are 
$\int d\omega\,\omega\,S(\omega)=0$ and 
$\int d\omega\,\omega^2\,S(\omega)=-\lambda_1/3$ \cite{Neubert:1993ch}.

It is appropriate at this point to relate our subleading shape functions to 
those defined by Bauer et al.\ \cite{Bauer:2001mh,Bauer:2002yu}. Using the 
notations of their second paper, we find $S(\omega) = f(-\omega)$ at leading
order, and
\begin{equation}
   s(\omega) = \frac{t(-\omega)}{2} \,, \quad
   t(\omega) = - h_1(-\omega) \,, \quad
   u(\omega) = - G_2(-\omega) \,, \quad
   v(\omega) = - H_2(-\omega)
\end{equation}
for the subleading functions. Their first paper uses a dimensionless momentum 
variable, so the relations are $S(\omega)=f(-\omega/m_b)/m_b$ etc. (In 
comparing, one must take note of the fact that we use a different sign 
convention for the Levi-Civita tensor. Also, we believe there are typographic 
errors in the definitions of the functions $g_2$ and $h_2$ in 
\cite{Bauer:2001mh}.)

Comparing our result (\ref{final}) for the hadronic tensor with the 
corresponding expressions obtained in \cite{Bauer:2001mh,Bauer:2002yu} we find 
some sources of disagreement. As mentioned earlier, the function $u(\omega)$ 
is divided by $\bar n\cdot p$, whereas Bauer et al.\ state that it enters in 
the universal combination $S(\omega)+u(\omega)/m_b$ with the leading shape 
function. Also, it appears to us that the term proportional to the trace 
$T_3$, which adds a contribution to the last structure function in 
(\ref{Wsl}), is missed by these authors.

\section{Contributions from four-quark operators}
\label{sec:4quark}

The last diagram in Figure~\ref{fig:graphs} shows a tree-level contribution to 
the hadronic tensor involving two insertions of the subleading SCET Lagrangian 
${\cal L}_{\xi q}^{(1)}$. The exchange of a hard-collinear gluon implies that
this graph is of order $g^2$, and so it vanishes in the limit where 
$\alpha_s$ is set to zero. The fact that the suppression factor is 
$\pi\alpha_s$ instead of $\alpha_s/\pi$ reflects the phase-space enhancement 
of four-quark tree-level graphs compared with loop diagrams 
\cite{Neubert:1996we}. One might therefore expect that the four-quark 
contribution is numerically as important as the other tree-level subleading
shape-function contributions.

Applying the partial-fraction identity (\ref{partialfractions}) twice, we find 
that the resulting contribution to the correlator $T_{ij}^{(2)}$ can be 
written as in (\ref{Tresult}), adding a fifth operator to the sum. It reads
\begin{equation}\label{O5}
   O_5(x_-) = \frac{\pi\alpha_s}{\bar n\cdot p}
   \int\limits_0^{\bar n\cdot x/2}\!dt_1
   \int\limits_{t_1}^{\bar n\cdot x/2}\!dt_2
   \left( \bar h S\right)_{0} \Gamma_i\,\nslash\gamma_\rho^\perp\,t_a
   \left( S^\dagger q \right)_{t_1 n}
   \left( \bar q S\right)_{t_2 n} \gamma_\perp^\rho\nslash\,\Gamma_j\,t_a
   \left( S^\dagger h \right)_{x_-} ,
\end{equation}
where $t_a$ are the generators of color SU($N_c$). Note that the field 
insertions are ordered according to ``light-cone time'' $x_-$, just as they 
appear in the Feynman diagram in Figure~\ref{fig:graphs}. This is a general 
result. Because the minus components $\bar n\cdot p_{hc}$ of hard-collinear 
momenta are large, of order $m_b$, hard-collinear fields always propagate 
forward in light-cone time. Turning, for instance, a forward-moving 
hard-collinear quark into a backward-moving hard-collinear anti-quark would 
require a hard quantum fluctuation, which is already integrated out in SCET. 
As a result, Feynman amplitudes in SCET are ordered with respect to light-cone 
time, and that ordering is preserved in the matching onto HQET. This 
discussion explains why all our operators have the property that the 
coordinates $z_-$ of soft fields range from 0 to $x_-$ in an ordered fashion. 
The results can therefore always be expressed in terms of bi-local operators 
depending only on 0 and $x_-$. However, at present we cannot exclude the 
possibility of non-trivial weight functions under these integrals, which could 
arise at higher orders in perturbation theory. If present, they may require a 
generalization of our definitions of subleading shape functions.

Returning to the case of the four-quark operator in (\ref{O5}), we note that 
its contribution vanishes in the vacuum-insertion approximation due to the 
color-octet structure of the heavy-light quark bilinears. While this 
approximation is admittedly naive, phenomenological evidence based on studies 
of $B$-meson lifetimes \cite{Neubert:1996we,Bigi:1994wa} and theoretical 
arguments based on lattice calculations 
\cite{DiPierro:1998ty,Becirevic:2001fy} and QCD sum rules \cite{Baek:1998vk}
support the notion that matrix elements which vanish in the vacuum-insertion 
approximation are numerically suppressed, typically by an order of magnitude. 
It is thus very unlikely that the four-quark operator in (\ref{O5}) could give 
a larger contribution than loop-suppressed $O(\alpha_s)$ corrections, which we 
have neglected. Note also that the contribution of the corresponding local 
operator to total inclusive rates, which is what remains when the 
corresponding shape functions are integrated over a sufficiently large domain, 
is bound to be tiny. The effect is of order $(\Lambda_{\rm QCD}/m_b)^3$, and 
it is $\alpha_s$-suppressed with respect to a four-quark contribution 
discussed by Voloshin \cite{Voloshin:2001xi}, whose effect on the total decay 
rate is believed to be at most 3\%. 

Nevertheless, it is interesting to study the structure of the operator
$O_5(x_-)$ in more detail and define corresponding subleading shape functions. 
The decomposition (\ref{HQETdecomp}) implies that the Dirac structure can be
rearranged in the form (omitting factors of $S$, $S^\dagger$ and color indices 
for simplicity)
\begin{equation}\label{step1}
   \Gamma_i\,\nslash\gamma_\rho^\perp\,q\,
   \bar q\,\gamma_\perp^\rho\nslash\,\Gamma_j
   = - \frac12 \left( \begin{array}{c} \bm{1} \\ \nslash\gamma_5 \\
   - \gamma_\perp^\mu\gamma_5 \end{array} \right)
   \bar q\,\gamma_\perp^\rho\nslash\,\Gamma_j\,\frac{1+\vslash}{2}
   \left( \begin{array}{c} \bm{1} \\ (\nslash-\vslash)\gamma_5 \\
   \gamma_\mu^\perp\gamma_5 \end{array} \right)
   \Gamma_i\,\nslash\gamma_\rho^\perp\,q \,,
\end{equation}
where the notation implies a sum over the three rows in the equation. In the
next step, we use that between $\nslash\dots\nslash$ any Dirac matrix can be
decomposed as
\begin{equation}
   \Gamma \to \frac14\,\mbox{tr}(\Gamma\,\nslash)\,\frac{\nbslash}{2}
   - \frac14\,\mbox{tr}(\Gamma\,\nslash\gamma_5)\,
   \frac{\nbslash}{2}\,\gamma_5
   - \frac14\,\mbox{tr}(\Gamma\,\nslash\gamma_\perp^\alpha)\,
   \frac{\nbslash}{2}\,\gamma_\alpha^\perp \,.
\end{equation}
Only the first two terms survive after contraction of the index $\rho$ in 
(\ref{step1}). At this stage we are left with four-quark operators with Dirac 
structures
\begin{equation}
   \bar h\,h\,\bar q\,\nslash\,(\gamma_5)\,q \,, \qquad
   \bar h\,\nslash\gamma_5\,h\,\bar q\,\nslash\,(\gamma_5)\,q \,, \qquad
   \bar h\,\gamma_\perp^\mu\gamma_5\,h\,\bar q\,\nslash\,(\gamma_5)\,q \,,
\end{equation}
where $(\gamma_5)$ means $\bm{1}$ or $\gamma_5$. Lorentz and parity invariance 
imply that only the two Lorentz-scalar structures 
$\bar h\,h\,\bar q\,\nslash\,q$ and 
$\bar h\,\nslash\gamma_5\,h\,\bar q\,\nslash\gamma_5\,q$ can have non-zero 
forward matrix elements between $B$-meson states. Putting back color factors 
and soft Wilson lines, we define the corresponding subleading shape functions
\begin{eqnarray}\label{4qdef}
   2(-i)^2 &&\hspace{-0.6cm} \int\limits_0^{\bar n\cdot x/2}\!dt_1
   \int\limits_{t_1}^{\bar n\cdot x/2}\!dt_2\,
   \langle \big[ \left( \bar h S\right)_{0} t_a \big]_k 
   \big[ t_a \left( S^\dagger h \right)_{x_-} \big]_l
   \big[ \left( \bar q S\right)_{t_2 n} \big]_l\,\nslash
   \big[ \left( S^\dagger q \right)_{t_1 n} \big]_k \rangle \nonumber\\
   &=& \int d\omega\,e^{-\frac{i}{2}\omega\bar n\cdot x}\,f_u(\omega) \,,
    \nonumber\\
   2(-i)^2 &&\hspace{-0.6cm} \int\limits_0^{\bar n\cdot x/2}\!dt_1
   \int\limits_{t_1}^{\bar n\cdot x/2}\!dt_2\,
   \langle \big[ \left( \bar h S\right)_{0} t_a \big]_k\,\nslash\gamma_5
   \big[ t_a \left( S^\dagger h \right)_{x_-} \big]_l
   \big[ \left( \bar q S\right)_{t_2 n} \big]_l\,\nslash\gamma_5
   \big[ \left( S^\dagger q \right)_{t_1 n} \big]_k \rangle \nonumber\\
   &=& \int d\omega\,e^{-\frac{i}{2}\omega\bar n\cdot x}\,f_v(\omega) \,,
\end{eqnarray}
where $k$, $l$ are color indices. Using the same notation as in (\ref{final}), 
the corresponding contributions to the hadronic tensor are given by
\begin{equation}
   W_{ij}^{(2)} \Big|_{\rm 4q}
   = - \pi\alpha_s \int d\omega\,\delta(n\cdot p+\omega)
   \left[ \frac{f_u(\omega)}{\bar n\cdot p}\,T_1
   + \frac{f_v(\omega)}{\bar n\cdot p}\,T_4 \right] ,
\end{equation}
with the same traces as defined in (\ref{traces}). As far as the spin structure
is concerned, the four-quark contributions can thus be absorbed into a 
redefinition of the subleading shape functions $u$ and $v$, namely
\begin{equation}\label{tildeuv}
   \tilde u(\omega)\equiv u(\omega) - \pi\alpha_s\,f_u(\omega) \,, \qquad
   \tilde v(\omega)\equiv v(\omega) + \pi\alpha_s\,f_v(\omega) \,.
\end{equation}
Note that the definitions (\ref{4qdef}) imply that the normalization integrals 
as well as the first moments of the functions $f_u(\omega)$ and $f_v(\omega)$ 
vanish (because the integration domain is of second order in $\bar n\cdot x$), 
and therefore the new functions $\tilde u$ and $\tilde v$ have the same first 
moments as the original ones, see (\ref{moments}). 

While the shape functions $S$, $s$, $t$, $u$, $v$ are expected to be identical 
for charged and neutral $B$ mesons up to tiny isospin-breaking corrections, 
this is no longer the case for the four-quark shape functions $f_u$ and $f_v$. 
The values of these functions will depend crucially on whether the light-quark 
flavor $q$ in the four-quark operators matches that of the $B$-meson spectator 
quark. In the semileptonic decay $\bar B\to X_u\,l^-\bar\nu$, the difference 
between the subleading shape functions $f_u$ and $f_v$ for $B^-$ and 
$\bar B^0$ mesons is likely to be one of the dominant sources of 
isospin-breaking effects on the decay distributions.

\section{Applications}

We are now ready to study the phenomenological implications of our results. We 
absorb the contributions from four-quark operators into the functions 
$\tilde u(\omega)$ and $\tilde v(\omega)$ defined in (\ref{tildeuv}). The 
moment constraints on the shape functions can be summarized by their 
expansions in distributions \cite{Neubert:1993ch}, which read
\begin{eqnarray}\label{distr}
   S(\omega) &=& \delta(\omega) - \frac{\lambda_1}{6}\,\delta''(\omega)
    + \dots \,, \quad
    s(\omega) = \frac{\lambda_1+3\lambda_2}{2}\,\delta'(\omega) + \dots \,, \\
   t(\omega) &=& \lambda_2\,\delta'(\omega) + \dots \,, \quad
    \tilde u(\omega) = - \frac{2\lambda_1}{3}\,\delta'(\omega) + \dots \,, 
    \quad
    \tilde v(\omega) = - \lambda_2\,\delta'(\omega) + \dots \,. \nonumber
\end{eqnarray}
These expressions allow us to test our results against existing predictions 
for inclusive spectra obtained using a conventional heavy-quark expansion. 

The analytic properties of the shape functions are such that they have support 
for $-\infty<\omega<\bar\Lambda_\infty$, where $\bar\Lambda_\infty$ is the 
asymptotic value of the mass difference $(m_B-m_b)_{m_b\to\infty}$ in the 
heavy-quark limit. This parameter differs from the physical value of 
$\bar\Lambda$ by power-suppressed terms, 
\begin{equation}
   \bar\Lambda\equiv m_B-m_b = \bar\Lambda_\infty
   - \frac{\lambda_1+3\lambda_2}{2m_b} + \dots \,.
\end{equation}
We would like the support in $\omega$ to extend over the physical interval
$-\infty<\omega<\bar\Lambda$, since this will ensure that the kinematic 
boundaries for decay distributions take their physical values set by the true 
$B$-meson mass. This can be achieved by shifting the arguments of all shape 
functions by a small amount $\Delta\omega=\frac12(\lambda_1+3\lambda_2)/m_b$. 
For the subleading shape functions this changes nothing to the order we are 
working, since $t(\omega+\Delta\omega)=t(\omega)+\dots$ etc., where the dots
represent terms of higher order in $1/m_b$. For the leading-order shape 
function, however, this shift produces a new $1/m_b$ correction:
$S(\omega)=S(\omega+\Delta\omega)-\Delta\omega\,S'(\omega+\Delta\omega)+\dots$,
where the prime denotes a derivative with respect to the argument. Using the 
fact that $S(\omega)$ and $s(\omega)/m_b$ always appear together, we can 
absorb the extra term into a redefinition of the subleading-shape function 
$s$, defining a new function 
\begin{equation}
   s_0(\omega)\equiv s(\omega) - \frac12(\lambda_1+3\lambda_2)\,S'(\omega) \,.
\end{equation}
From (\ref{moments}), it follows that the first moment of the function $s_0$ 
vanishes. In terms of these definitions,
\begin{equation}
   \tilde S(\omega+\Delta\omega)
   \equiv S(\omega+\Delta\omega) + \frac{s_0(\omega+\Delta\omega)}{m_b}
   = S(\omega) + \frac{s(\omega)}{m_b} + \dots \,.
\end{equation}
In other words, our expressions for the hadronic tensors in (\ref{Wsl}) and 
(\ref{Wgamma}) remain valid when in all shape functions the argument is 
shifted from $\omega$ to $\omega+\Delta\omega$, except that the subleading 
shape function $s$ must be replaced with the redefined function $s_0$, whose 
norm and first moment vanish. Once this is done, the integrals over $\omega$ 
extend from $-\infty$ up to the physical value of $\bar\Lambda$.

It is now convenient to introduce a new variable 
$P_+=n\cdot p+\bar\Lambda=m_B-n\cdot q=E_X-|\bm{P}_X|$, which is the plus 
component of the total momentum of the final-state hadronic jet, and to 
express the shape functions as functions of 
$\hat\omega\equiv\bar\Lambda-\omega$ \cite{Bosch:2004th}. They have support in 
this variable for $0\le\hat\omega<\infty$. The $\delta$-functions in the 
tree-level expressions (\ref{Wsl}) and (\ref{Wgamma}) for the hadronic tensors 
set $\hat\omega=P_+$. We denote functions of $\hat\omega$ by a hat, e.g.\
$\hat S(\hat\omega)\equiv\tilde S(\bar\Lambda-\hat\omega+\Delta\omega)%
=\tilde S(\bar\Lambda_\infty-\hat\omega)$, 
$\hat t(\hat\omega)\equiv t(\bar\Lambda_\infty-\hat\omega)$, and similarly for 
the other functions. In the equations below, $\bar\Lambda$ always refers to 
the physical parameter defined with the true $B$-meson mass.

We start by presenting the triple differential rate for the inclusive 
semileptonic decay $\bar B\to X_u\,l^-\bar\nu$ in terms of the kinematic 
variables
\begin{equation}
   P_+ \,, \qquad
   y = \frac{\bar n\cdot p}{m_b} = 1 - \frac{\bar n\cdot q}{m_b} \,, \qquad
   \bar x = 1 - \frac{2E_l}{m_b} \,.
\end{equation}
Using the expression for the hadronic tensor in (\ref{Wsl}) combined with the
rate equations derived in \cite{DeFazio:1999sv}, we obtain
\begin{eqnarray}\label{d3G}
   \frac{1}{\Gamma}\,\frac{d^3\Gamma}{d\bar x\,dy\,dP_+}
   &=& 12(y-\bar x)\,\Bigg\{
    (1-y+\bar x)\!\left[ \left( 1 + \frac{2(\bar\Lambda-P_+)}{m_b} \right)
    \hat S(P_+) + \frac{\hat t(P_+)}{m_b} 
    + \frac{\hat u(P_+) - \hat v(P_+)}{y\,m_b} \right] \nonumber\\
   &&\mbox{}+ \left( 1 - \frac{2\bar x}{y} \right)
    \frac{\bar\Lambda-P_+}{m_b}\,\hat S(P_+)
    + \frac{2\bar x(1-2y)}{y^2}\,\frac{\hat t(P_+)}{m_b}
    + \frac{2\bar x}{y^2}\,\frac{\hat v(P_+)}{m_b} \Bigg\} \,.
\end{eqnarray}
This expression is exact at tree level and to order $\Lambda_{\rm QCD}/m_b$ in 
the heavy-quark expansion. The phase space for these variables is such that 
\begin{equation}
   0\le P_+\le m_B-2E_l = m_b\,\bar x + \bar\Lambda \,, \qquad
   \frac{P_+ - \bar\Lambda}{m_b}\le\bar x\le y\le 1 \,, 
\end{equation}
and the hadronic tensor has support only for $y\ge 0$. Above we assume that 
$\bar x$ and $y$ are of $O(1)$. The collinear expansion breaks down in the 
region of small $y$. A restriction to small $\bar x$ is allowed and 
corresponds to the so-called ``endpoint region'' of the charged-lepton energy 
spectrum. In this case, however, it is necessary for consistency to expand the 
expression above in powers of $\bar x=O(\Lambda_{\rm QCD}/m_b)$.

As a first application, we integrate (\ref{d3G}) over the lepton energy, 
finding
\begin{eqnarray}
   \frac{1}{\Gamma}\,\frac{d^2\Gamma}{dy\,dP_+}
   &=& 2y^2(3-2y) \left[ \left( 1 + \frac{2(\bar\Lambda-P_+)}{m_b} \right)
    \hat S(P_+) + \frac{\hat t(P_+)}{m_b} + \frac{\hat u(P_+)}{y\,m_b} \right] 
    \nonumber\\
   &&\mbox{}+ 2y(6-5y)\,\frac{\bar\Lambda-P_+}{m_b}\,\hat S(P_+)
    + 2y(1-2y)\,\frac{2\hat t(P_+) - \hat v(P_+)}{m_b} \,.
\end{eqnarray}
The variables $y$ and $P_+$ allow us to reconstruct any kinematic quantity
characterizing the final-state hadronic system, such as the total hadronic 
energy, $2E_X=P_+ +y m_b+\bar\Lambda$, or the invariant hadronic mass, 
$M_X^2=P_+(y m_b+\bar\Lambda)$. Of particular interest for a measurement of 
$|V_{ub}|$ are the spectra in the variables $P_+$ and $M_X$. For the 
distribution in $P_+$, we obtain
\begin{equation}\label{Ppl}
   \frac{1}{\Gamma}\,\frac{d\Gamma}{dP_+}
   = \left[ 1 + \frac{14(\bar\Lambda-P_+)}{3m_b} \right] \hat S(P_+)
   + \frac{\hat t(P_+) + 5\hat u(P_+) + \hat v(P_+)}{3m_b} \,.
\end{equation}
Next, denoting $s_H=M_X^2$ and $\Delta_s=s_H/m_B$, we find for the hadronic 
invariant mass spectrum
\begin{equation}\label{sH}
   \frac{1}{\Gamma}\,\frac{d\Gamma}{ds_H}
   = \frac{1}{m_B} \int\limits_{\Delta_s}^\infty \frac{dP_+}{P_+}\,
   f(P_+,\Delta_s/P_+) \,,
\end{equation}
where
\begin{eqnarray}
   f(P_+,r)
   &=& 2r^2(3-2r) \left[ \left( 1 + \frac{2(3\bar\Lambda-P_+)}{m_b} \right)
    \hat S(P_+) + \frac{\hat t(P_+)}{m_b} + \frac{\hat u(P_+)}{r\,m_b}
    \right] \nonumber\\
   &&\mbox{}- 2r \left[ (6-5r)\,\frac{P_+}{m_b}
    + 2r\,\frac{\bar\Lambda}{m_b} \right] \hat S(P_+)
    + 2r(1-2r)\,\frac{2\hat t(P_+) - \hat v(P_+)}{m_b} \,.
\end{eqnarray}
A comment is in order concerning the fact that the integration over $P_+$ in 
(\ref{sH}) is extended to infinity, while kinematically $P_+\le\sqrt{s_0}$. 
The point is that near its kinematic limit the momentum 
$P_+\sim\sqrt{m_B\Lambda_{\rm QCD}}$ is no longer of order 
$\Lambda_{\rm QCD}$, and hence the collinear expansion breaks down. However, 
the contribution to the rate resulting from the vicinity of this region is 
suppressed by more than a single power of $\Lambda_{\rm QCD}/m_b$ 
\cite{Bosch:2004th}. It is therefore required to set the upper limit to 
infinity so as to avoid spurious higher-order power corrections. Once this is 
done, it is theoretically consistent to treat $P_+$ as a quantity of order 
$\Lambda_{\rm QCD}$.

As a final application, we integrate (\ref{d3G}) over the hadronic variables 
$y$ and $P_+$ to obtain the charged-lepton energy distribution
\begin{equation}\label{El}
   \frac{1}{\Gamma}\,\frac{d\Gamma}{dE_l}
   = 4 \int\limits_0^{m_B-2E_l}
   \frac{d\hat\omega}{m_B-\hat\omega} \left[ \left(
   1 + \frac{6(\bar\Lambda-\hat\omega)}{m_b} \right) \hat S(\hat\omega)
   + \frac{\hat t(\hat\omega) + 3\hat u(\hat\omega)
           - 3\hat v(\hat\omega)}{m_b} \right] .
\end{equation}

For purposes of comparison with previous authors, we also give the result for
the dipole-operator contribution to the $\bar B\to X_s\gamma$ photon spectrum, 
stressing once again that this does {\em not\/} provide a complete description 
of all tree-level subleading shape-function effects. Taking into account 
that the photon spectrum is proportional to the expression in (\ref{Wgamma}) 
times a factor $E_\gamma^3$, we find
\begin{equation}\label{photon}
   \frac{1}{2\Gamma}\,\frac{d\Gamma}{dE_\gamma}
   = \left[ 1 + \frac{2(\bar\Lambda-P_+)}{m_b} \right] \hat S(P_+)
   + \frac{- \hat t(P_+) + \hat u(P_+) - \hat v(P_+)}{m_b} 
   + \dots \,,
\end{equation}
where $P_+=m_B-2E_\gamma$ in this case.

Despite the discrepancies mentioned earlier, we agree with \cite{Bauer:2001mh} 
on the result for the $\bar B\to X_s\gamma$ photon spectrum (having, however, 
added the new contribution from four-quark operators), and we also find 
complete agreement with a recent calculation of the hadronic invariant mass 
spectrum for semileptonic $\bar B\to X_u\,l^-\bar\nu$ decay by Luke et al.\ 
\cite{Burrell:2003cf}. However, we do not confirm the result for the 
charged-lepton energy spectrum obtained in \cite{Bauer:2002yu}, when converted 
to our notation. These authors have (apart from an overall missing factor of 
2) a coefficient of 2 instead of 
our 6 in front of the term proportional to $(\bar\Lambda-\hat\omega)$, and a 
coefficient of 1 instead of our 3 in front of $\hat u(\hat\omega)$. Our 
results can be tested against established formulae by using the moment 
expansions in (\ref{distr}). We find that (\ref{Ppl}), (\ref{El}), and 
(\ref{photon}) are consistent with expressions for the corresponding spectra 
derived in \cite{Bosch:2004bt}, \cite{Manohar:1993qn}, and 
\cite{Neubert:1993ch}, respectively. Amusingly, the result for the lepton 
spectrum found in \cite{Bauer:2002yu} is also consistent with the moment 
expansion. The reason is that this spectrum differs from our result by a term 
involving the combination $4(\omega/m_b)\,\tilde S(\omega)+2\tilde u(\omega)$, 
in which the $\delta'(\omega)$ terms from (\ref{distr}) cancel each other.

Of particular phenomenological importance are various fractions of all 
$\bar B\to X_u\,l^-\bar\nu$ decays that pass certain experimental cuts, which 
are chosen so as to eliminate (or reduce) the background from semileptonic
decays with charm hadrons in the final state. The most common cuts are
$E_l\ge E_0$ with $\Delta_E=m_B-2E_0\le m_D^2/m_B$ (``lepton endpoint''), 
$s_H\le s_0$ with $\Delta_M=s_0/m_B\le m_D^2/m_B$ (``low hadronic mass''), 
and $P_+\le\Delta_P$ with $\Delta_P\le m_D^2/m_B$ (``low hadronic plus 
momentum''). We define corresponding fractions $F_E(\Delta_E)$, 
$F_M(\Delta_M)$, and $F_P(\Delta_P)$. While the result for $F_P(\Delta_P)$ is 
simply an integral of the spectrum in (\ref{Ppl}) over the range 
$0\le P_+\le\Delta_P$, the expressions for the other two event fractions are 
more complicated, and we quote them here for completeness. We obtain
\begin{equation}\label{FE}
   F_E(\Delta_E) = \int\limits_0^{\Delta_E}\!d\hat\omega\,
    \frac{2(\Delta_E-\hat\omega)}{m_B-\hat\omega} \left[ \left(
    1 + \frac{6(\bar\Lambda-\hat\omega)}{m_b} \right) \hat S(\hat\omega)
    + \frac{\hat t(\hat\omega) + 3\hat u(\hat\omega) - 3\hat v(\hat\omega)}
           {m_b} \right] ,
\end{equation}
and
\begin{eqnarray}\label{FMinteg}
   F_M(\Delta_M) &=& \int\limits_0^{\Delta_M}\!dP_+\,
    \left[ \left( 1 + \frac{14(\bar\Lambda-P_+)}{3m_b} \right) \hat S(P_+)
    + \frac{\hat t(P_+) + 5\hat u(P_+) + \hat v(P_+)}{3m_b} \right]
    \nonumber\\
   &+& \int\limits_{\Delta_M}^\infty\!dP_+\,g(P_+,\Delta_M/P_+) \,,
\end{eqnarray}
where $\Delta_M=s_0/m_B$, and 
\begin{eqnarray}
   g(P_+,r)
   &=& r^3(2-r) \left[ \left( 1 + \frac{2(3\bar\Lambda-P_+)}{m_b} \right)
    \hat S(P_+) + \frac{\hat t(P_+)}{m_b} \right]\!
    - \frac{2r^2}{3}\!\left[ (9-5r) \frac{P_+}{m_b}
    + 2r\,\frac{\bar\Lambda}{m_b} \right] \hat S(P_+) \nonumber\\
   &&\mbox{}+ \frac{r^2}{3}\,(9-4r) \frac{\hat u(P_+)}{m_b}
    + \frac{r^2}{3}\,(3-4r)\,\frac{2\hat t(P_+) - \hat v(P_+)}{m_b} \,.
\end{eqnarray}

In order to illustrate the numerical impact of subleading shape functions, we 
focus on the single differential spectra in (\ref{Ppl}), (\ref{sH}), and
(\ref{El}) in semileptonic $B$ decay. For the leading shape function we use 
the parametrization
\begin{equation}\label{Smodel}
   \hat S_{\rm model}(\hat\omega) = \frac{b^b}{\Gamma(b)}\,
   \frac{\hat\omega^{b-1}}{\bar\Lambda^b}\,e^{-b\,\hat\omega/\bar\Lambda}
\end{equation}
with $\bar\Lambda=0.63$\,GeV and $b=2.93$, corresponding to the default choice 
in \cite{Bosch:2004th}. We employ two models for the subleading shape 
functions based on the moment expansion in (\ref{distr}), one where 
$\delta'(\omega)$ is replaced by a derivative 
$-\hat S_{\rm model}'(\hat\omega)$ of the leading shape function, and one 
where it is replaced by 
$[3(\bar\Lambda-\hat\omega)/\lambda_1]\,\hat S_{\rm model}(\hat\omega)$. Both 
models satisfy the moment constraints identically, while giving rather 
different shapes for the subleading structure functions. We take 
$\lambda_2=0.12$\,GeV$^2$, $m_b=4.65$\,GeV, and determine 
$\lambda_1=-0.41$\,GeV$^2$ from the second moment of the model function 
(\ref{Smodel}). 

\begin{figure}[t]
\begin{center}
\epsfig{file=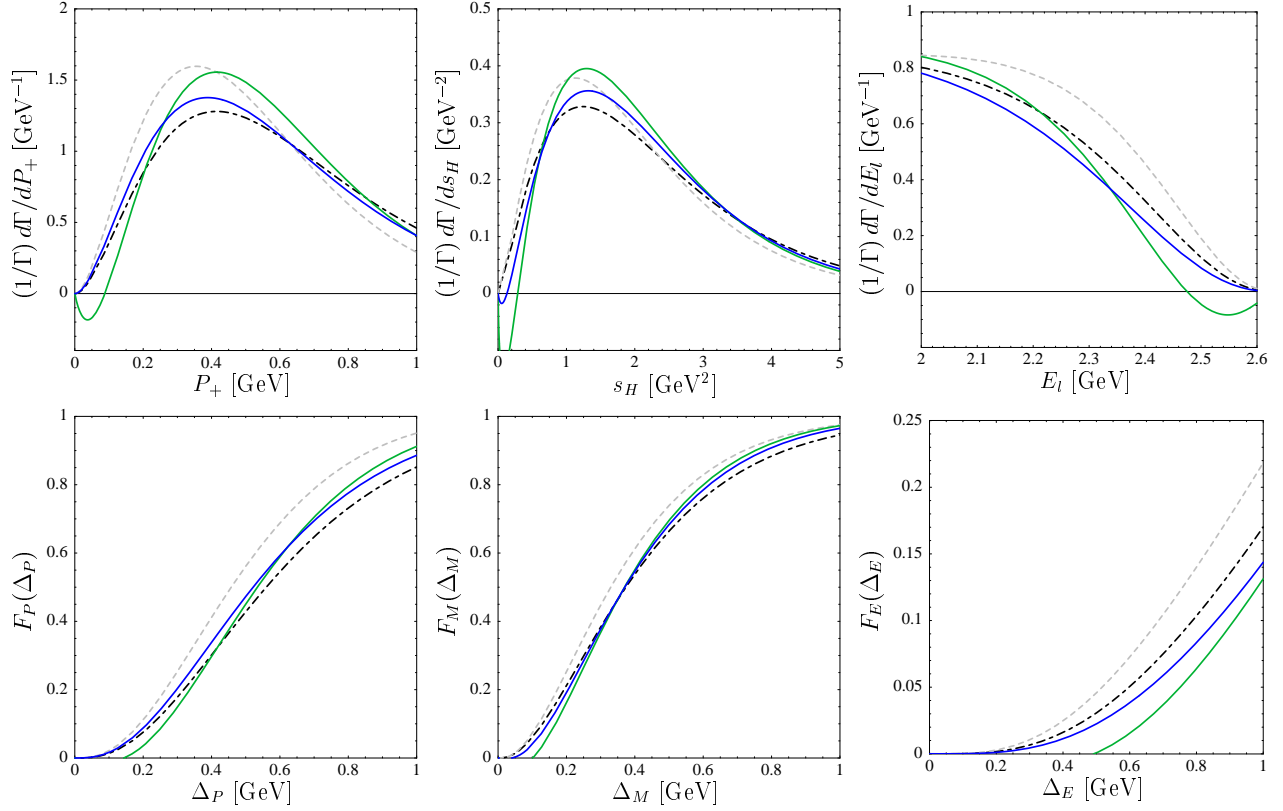,width=\textwidth}
\end{center}
\vspace{-0.2cm}
\centerline{\parbox{15cm}{\caption{\label{fig:spectra}
Predictions for the $P_+$, $s_H$, and $E_l$ spectra (upper plots) and event 
fractions (lower plots). The dash-dotted curves refer to the heavy-quark 
limit, while the solid lines include subleading shape-function effects. The 
gray dashed curves include only those power corrections proportional to the 
leading-order shape function.}}}
\end{figure}

Results for the three spectra and the corresponding integrated event fractions 
are presented in Figure~\ref{fig:spectra}. In each plot, the dashed-dotted 
line shows the leading term in the heavy-quark expansion, the dashed gray 
curve includes the effect of the ``kinematic'' power corrections proportional 
to $\hat S(\hat\omega)$, which are predicted model independently, and the two 
solid lines include the model predictions for the subleading shape-function 
effects (the lighter of the two lines corresponds to the first model). While 
the effects are clearly visible on the spectra themselves, the integrated rate 
fractions with a cut on $P_+$ or $s_H$ receive only moderate power 
corrections, which alter the leading-power results by typically 10\% or less 
in the region near the charm threshold (located at 
$m_D^2/m_B\approx 0.66$\,GeV). Note that the kinematic corrections alone 
overestimate the effect, especially for the case of the $P_+$ spectrum. 

The situation is rather different for the charged-lepton energy spectrum, 
where power corrections can have a drastic effect and tend to reduce the event 
fraction by a large amount. This effect can be traced to the particular weight 
function under the integral in (\ref{FE}) \cite{Neubert:2002yx}. Even for 
large values of $\hat\omega$, the effect of subleading shape functions does 
not diminish, because the linear term in the weight function picks out the 
first moments of the subleading shape functions. As a result, there is a 
significant theoretical uncertainty in the prediction for the event fraction 
$F_E(\Delta_E)$ near the charm threshold, which imposes limitations on the 
precision with which $|V_{ub}|$ can be determined from the charged-lepton 
endpoint region.

No such large effects are seen for the event fractions with a cut on $P_+$ or
hadronic invariant mass. With $\Delta_{P,M}\approx m_D^2/m_B$, about 65--85\% 
of all $\bar B\to X_u\,l^-\bar\nu$ events are retained with these cuts, and 
the corresponding fractions are enhanced further when radiative corrections 
are included \cite{Bosch:2004th}. Once the efficiency is so high, possible 
uncertainties due to four-quark operator contributions (``weak annihilation'') 
become negligible. Our analysis supports the study of theoretical 
uncertainties presented in \cite{Bosch:2004bt}, where the uncertainty due to 
power corrections in the value of $F_P(\Delta_P)$ near the charm threshold was 
estimated as 10\%. It provides further credibility to the strategy of 
obtaining highly efficient, precise measurements of $|V_{ub}|$ using 
experimental cuts on $P_+$ or $s_H$. We recall at this point that the $P_+$ 
spectrum offers the additional advantage that it can be directly related to 
the $\bar B\to X_s\gamma$ photon spectrum, cf.~(\ref{Ppl}) and (\ref{photon}), 
whereas a prediction of the hadronic invariant mass spectrum requires 
knowledge of the shape function over a range not accessible in radiative $B$ 
decay. Note also that the hadronic invariant mass distribution receives 
fractional power corrections of order $(\Lambda_{\rm QCD}/m_b)^\gamma$ with 
$\gamma\approx 1.35$ from a region of phase space where the collinear 
expansion breaks down \cite{Bosch:2004th}. While these effects are formally 
beyond the accuracy of the present work, they introduce an additional 
theoretical uncertainty that is difficult to quantify (unless they can be 
shown to cancel).

The smallness of subleading shape-function effects for some observables can be 
understood based on the moment expansions in (\ref{distr}). For instance, in 
the $P_+$ spectrum (\ref{Ppl}) the first moments of the functions $\hat t$ and 
$\hat v$ cancel in the combination $\hat t+\hat v$, and the first moment of 
$\frac53\,\hat u$ cancels against that of 
$\frac{10}{3}\,(\bar\Lambda-P_+)\,\hat S$, leaving a small residual 
term $-\frac49\,\lambda_1\,\delta'(P_+ -\bar\Lambda)$. In weighted integrals
over the $P_+$ spectrum, this can give rise to small second-order power 
corrections. Higher moments may spoil these cancellations, but their effects 
are only of order $(\Lambda_{\rm QCD}/m_b)^3$ or higher when the spectrum is 
integrated over a sufficiently large domain. Similar cancellations occur for 
the hadronic invariant mass spectrum (\ref{sH}), and for the photon spectrum 
(\ref{photon}), for which the first moments cancel entirely. In all three 
cases, we have checked that taking a mix of the two models for the subleading 
shape functions, in which different higher moments prevent perfect 
cancellations, does not change the results by a significant amount. The case 
of the charged-lepton spectrum (\ref{El}) is again different. The first moment 
of $3\hat u$ cancels against that of $6(\bar\Lambda-\hat\omega)\,\hat S$. 
However, the sum $\hat t-3\hat v$ has a rather large first moment given by 
$-4\lambda_2\,\delta'(\hat\omega-\bar\Lambda)$, which in conjunction with the 
linear weight factor in (\ref{FE}) gives rise to a significant power 
correction to the event fraction $F_E$ 
\cite{Leibovich:2002ys,Bauer:2002yu,Neubert:2002yx}.

\section{Conclusions}

With the development of soft-collinear effective theory (SCET), theoretical 
predictions for inclusive $B$-meson decay distributions have recently received 
renewed attention. While these processes had been studied extensively during 
the past decade, the ever increasing precision of the measurements at the $B$ 
factories requires a new level of accuracy. This is an area where SCET is not 
only a useful conceptual tool, but where it provides concrete means of pushing 
the limits of theoretical calculations.

Earlier this year, two groups have performed systematic analyses of 
short-distance corrections in the semileptonic decay 
$\bar B\to X_u\,l^-\,\bar\nu$ \cite{Bosch:2004th,Bauer:2003pi}
and radiative decay $\bar B\to X_s\gamma$ \cite{Neubert:2004dd}, including a 
complete scale separation and resummation of Sudakov logarithms. As a result, 
the behavior of the leading shape function under renormalization is now well 
understood. This leaves power corrections to the heavy-quark limit as the 
principal source of theoretical uncertainties.

In the present work, we have used the formalism of SCET to perform a 
systematic study of such power-suppressed effects. At tree level, the results 
can be expressed in terms of a set of subleading shape functions defined via 
the Fourier transforms of forward matrix elements of bi-local light-cone 
operators in heavy-quark effective theory. We have identified a new 
contribution arising from four-quark operators, which was not considered 
previously. We have shown that, when shape functions appearing in 
process-independent combinations are combined into single functions, then a 
total of three subleading shape functions are required to describe arbitrary 
current-induced decay distributions of $B$ mesons into light final-state 
particles. While subleading shape-function effects had been studied in the 
past, our results do not agree with those found in the original papers 
\cite{Bauer:2001mh,Bauer:2002yu}.

In the last part of this work, we have presented analytical expressions for a 
variety of distributions in $\bar B\to X_u\,l^-\bar\nu$ decay, which can be 
used directly for the analysis of experimental data. We have also given a 
formula for the triple differential rate, which allows for arbitrary cuts on 
kinematic variables. While this concludes the problem of tree-level power 
corrections in semileptonic decay, we have stressed that no complete 
(tree-level) analysis of power-suppressed corrections to the 
$\bar B\to X_s\gamma$ decay exists to date. The formalism developed in this 
work can, however, readily be extended to this case.

%\vspace{0.5cm}\noindent
\newpage
{\em Acknowledgments:\/}
We are grateful to Martin Beneke, Masahiro Morii, and Ben Pecjak for useful 
comments. S.W.B.\ would like to thank Daniel Wyler at the University of Zurich 
for hospitality during the final stage of this work. This research was 
supported by the National Science Foundation under Grant PHY-0355005. The 
research of M.N.\ is also supported in part by the Department of Energy under 
Grant DE-FG02-90ER40542.

\end{document}